\documentclass[twocolumn]{aastex631}

\received{October 11, 2023}
\revised{December 8, 2023}
\accepted{December 10, 2023}

\submitjournal{ApJL}

\renewcommand{\deg}{^\circ}
\newcommand{\h}{^\mathrm{h}}
\newcommand{\m}{^\mathrm{m}}
\newcommand{\s}{^\mathrm{s}}

\newcommand{\au}{\,\mathrm{au}}
\newcommand{\km}{\,\mathrm{km}}

\newcommand{\cm}{\,\mathrm{cm}}

\newcommand{\Myr}{\,\mathrm{Myr}}
\newcommand{\kyr}{\,\mathrm{kyr}}

\newcommand{\K}{\,\mathrm{K}}
\newcommand{\J}{\,\mathrm{J}}
\newcommand{\N}{\,\mathrm{N}}

\newcommand{\g}{\,\mathrm{g}}

\newcommand{\W}{\,\mathrm{W}}

\begin{document}

\title{Tidal disruption of near-Earth asteroids during close encounters with terrestrial planets}

\correspondingauthor{Mikael Granvik}
\email{mgranvik@iki.fi}

\author[0000-0002-5624-1888]{Mikael Granvik}
\affiliation{Asteroid Engineering Laboratory, Lule\aa{} University of Technology, Box 848, Kiruna, Sweden}
\affiliation{Department of Physics, P.O. Box 64, 00014 University of Helsinki, Finland}

\author[0000-0002-0906-1761]{Kevin J. Walsh}
\affiliation{Southwest Research Institute, 1050 Walnut St, Suite 300, Boulder, CO 80302, U.S.A.}



\begin{abstract}

Numerical modeling has long suggested that gravitationally-bound (or so-called rubble-pile) near-Earth asteroids (NEAs) can be destroyed by tidal forces during close and slow encounters with terrestrial planets. However, tidal disruptions of NEAs have never been directly observed nor have they been directly attributed to any families of NEAs. Here we show population-level evidence for the tidal disruption of NEAs during close encounters with the Earth and Venus. Debiased model distributions of NEA orbits and absolute magnitudes based on observations by the Catalina Sky Survey during 2005--2012 underpredict the number of NEAs with perihelion distances coinciding with the semimajor axes of Venus and the Earth. A detailed analysis of the orbital distributions of the excess NEAs shows that their characteristics agree with the prediction for tidal disruptions, and they cannot be explained by observational selection effects or orbital dynamics. Accounting for tidal disruptions in evolutionary models of the NEA population partly bridges the gap between the predicted rate of impacts by asteroids with diameters of tens of meters and observed statistics of fireballs in the same size range.

\end{abstract}



\section{Introduction} \label{sec:intro}

The disruption of comet Shoemaker-Levy 9 during a close passage of Jupiter illuminated the weak gravitationally-bound interior structure of small bodies now often referred to as rubble piles \citep{2002aste.book..501R,2018ARA&A..56..593W}. The details of the disruption, the size and spacing of the fragment train in particular, provided significant leverage for models of tidal disruption to constrain the comet's original size and density \citep{1994Natur.370..120A}. Numerical models of the tidal disruption of rubble piles continued to gain capability and sophistication, and have since surveyed possible outcomes of encounters of rubble-pile asteroids with terrestrial planets. These simulations have accounted for minimum encounter distance and encounter speed, as well as the progenitors shape and spin \citep{1998Icar..134...47R}, and shear strength by way of surface friction \citep{2020A&A...640A.102Z}.

Tidal disruption has been pointed to as a likely mechanism in re-shaping some enigmatic asteroids, where the shape of asteroid (1620) Geographos is a primary suspect \citep{1998Icar..134...47R}. Similarly, \citet{2014Icar..238..156S} postulated that tidal disruption during a close Earth encounter was a source for near-Earth-object (NEO) families and estimated the orbital evolution of these families over time to understand why none have been identified to date \citep[see, e.g.,][]{2012Icar..220.1050S}. They concluded that the decoherence time of NEO families is too short compared to the frequency of tidal-disruption events to allow NEO families to be identified at any given time. There has thus never been any observational evidence suggesting that the tidal disruption of asteroids during close planetary encounters would be an important aspect of asteroid evolution in the inner Solar System.

Meanwhile, the discrepancy between the observed \citep{2013Natur.503..238B} and predicted \citep{2015Icar..257..302H,2021Icar..36514452H} rate of small asteroid or meteoroid impacts with the Earth has not been conclusively solved to date. The explanations range from extrinsic reasons such as systematic errors in the analysis of optical impact flashes to intrinsic reasons such as the asteroid albedo changing with diameter. Detailed analysis by \citet{boslough2015} reduced the discrepancy but a factor of few still remains. An excess of low-inclination Aten asteroids (semimajor axis $a<a_{\rm Earth}$ and aphelion distance $Q>q_{\rm Earth}$, where $q_{\rm Earth}$ is the perihelion distance of the Earth) has also been reported but conclusive evidence for its origin has so far been lacking \citep{2012ApJ...752..110M,2013ApJ...767L..18G}.

The actual population of objects on near-Earth orbits is vastly better constrained than just a decade ago owing to numerous surveys with complementary approaches and long timelines of operation such as the Catalina Sky Survey (CSS). These data provide powerful constraints on numerical models describing the debiased distribution of orbital elements and absolute magnitudes of NEOs \citep{2016Natur.530..303G,2018Icar..312..181G,2023AJ....166...55N}. While nominally simulating the entire near-Earth population in a steady-state scenario, one outcome focused primarily on the discrepancy between observed and predicted number of asteroids at small perihelion distances. After carefully making sure that the discrepancy is statistically significant and that it is not caused by errors in any aspects of the modeling, \citet{2016Natur.530..303G} concluded that asteroids are essentially completely destroyed---hence the term super-catastrophic disruption---close to the Sun but at distances that are nontrivial to explain. The finding has later been confirmed \citep{2018Icar..312..181G,2023AJ....166...55N}.

The fidelity of the latest NEO population models allow for a direct comparison with observed Earth and/or Venus crossing populations to search of over-predictions or under-predictions that could be related to tidal disruption. Here we take a closer look at the region in orbital-element space surrounding the orbits of Venus and Earth, and compare the observed population to theoretical predictions for tidal disruptions during close encounters with these planets.

\section{Data and methods}

Let us first summarize the data and methods that underlie the debiased model of NEO orbits and absolute magnitudes. The choice to focus on the model by \citet{2016Natur.530..303G} rather than a more recent model, such as \citet{2018Icar..312..181G} or \citet{2023AJ....166...55N}, is that the former was extensively scrutinized to give credibility to the discovery of super-catastrophic disruptions by ruling out all possible issues with the modeling approach. In addition, \citet{2016Natur.530..303G} model super-catastrophic disruption explicitly as a cut-off affecting individual test asteroids during the orbital integrations rather than a mathematical penalty function affecting the resulting orbit distribution, and is therefore conceptually intuitive and easy to understand. Finally, all of the aforementioned models are based on the same observational data set from CSS, and have been shown to be in general agreement with each other.

The fundamental equation solved when constructing an NEO population model is
\begin{eqnarray}
  n(a,e,i,H) = \epsilon(a,e,i,H)\,\times\,M(a,e,i,H) = \nonumber \\ 
   \epsilon(a,e,i,H)\,\times\,\sum_{s=1}^{N_{\rm ER}} N_s(H)\,R_s(a,e,i)\,, \label{eq:impeff}
\end{eqnarray}
where $n(a,e,i,H)$ is the number distribution of NEOs detected by a survey during some time interval in the space of orbital elements (semimajor axis $a$, eccentricity $e$, and inclination $i$) and absolute magnitude ($H$), $\epsilon(a,e,i,H)$ is the so-called bias correction function which provides an absolutely-calibrated estimate for the number of NEOs that should be detected by the same survey during the same time interval \citep{2016Icar..266..173J}, and $M(a,e,i,H)$ is the debiased model that we want to derive. To constrain the model in a physically-meaningful way, we separate the debiased model into its components: $N_{\rm ER}$ is the number of escape regions (ER) from which asteroids and comets enter the NEO region (also sometimes called source regions) considered in the model, and $N_s(H)$ and $R_s(a,e,i)$ are the $H$-frequency distribution and the normalized, steady-state orbit distribution, respectively, for NEOs originating in ER $s$. The steady-state orbital distributions, $R_s(a,e,i)$, are estimated numerically by following the orbital evolution of numerous test bodies from the main asteroid belt and cometary reservoirs into the NEO region, and recording the time that the test bodies spend in various parts of the ($a,e,i$) space in the NEO region \citep{2016Natur.530..303G,2017A&A...598A..52G,2018Icar..312..181G}.

\citet{2016Natur.530..303G} used a parameterization for the differential $H$ distribution that allows for a smooth, second-degree variation of the slope:
\begin{eqnarray}
  N_s(H) = \nonumber \\ 
  N_s(H; N_{0, s}, \alpha_{{\rm min},s}, H_{{\rm min},s}, c_s) = \nonumber \\ 
  N_{0,s}\,10^{\int_{H_0}^{H}\left[\alpha_{{\rm min},s} + c_s(H'-H_{{\rm min},s})^2\right]\,dH'} = \nonumber \\
  N_{0,s}\,10^{\alpha_{{\rm min},s}(H-H_0) + \frac{c_s}{3}\left[(H-H_{{\rm min},s})^3 - (H_0-H_{{\rm min},s})^3\right]} \,. \label{eq:hdistr}
\end{eqnarray}
The model by \citet{2016Natur.530..303G} is calibrated with CSS's detections of NEOs with $17<H<25$ obtained during 2005--2012. The free parameters fitted with a simplex method are those describing the $H$ distributions, that is, $N_{0, s}$, $\alpha_{{\rm min},s}$, $H_{{\rm min},s}$, and $c_s$.

There are thus no knobs that could be turned in the presented methodology to either produce or get away with features in the resulting debiased orbit and absolute-magnitude distribution, $M(a,e,i,H)$, other than by introducing new escape regions or source regions for NEOs, or otherwise modify the input orbit distributions.

\section{Results and discussion}

\begin{figure}[h!]
  \centering
  \includegraphics[width=0.45\textwidth]{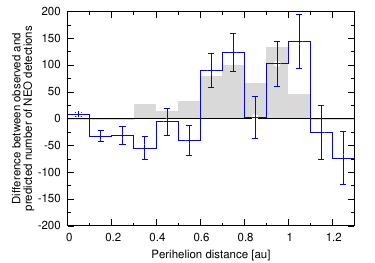}
  \caption{The difference between observed and predicted number of NEO detections by CSS during the years 2005--2012 as a function of perihelion distance $q$ (blue line). The model prediction assumes a super-catastrophic disruption when $q\sim0.076\au$ \protect \citep{2016Natur.530..303G}. The observed population is substantially larger than the predicted population for $q \sim a_{\rm Venus}\sim0.7\au$ and $q \sim a_{\rm Earth}\sim1\au$. The difference cannot be explained by selection effects or orbital dynamics. The gray histogram shows an arbitrarily-normalized distribution of the perihelion distances of synthetic gravitational aggregates that in numerical simulations have undergone B-type tidal disruptions during encounters with the Earth or Venus.}
  \label{fig:absolutediff_vs_q_17H25}
\end{figure}

\citet{2016Natur.530..303G} found that, by assuming a complete, instantaneous destruction of asteroids at an average perihelion distance $q=0.076\au$, the model could reproduce the observed perihelion distances $q\lesssim0.6\au$ significantly more accurately than without assuming a destruction (see their Fig.~1). Note, however, that the rather simplistic disruption model, which averages over all orbits, taxonomic types, and sizes, and is agnostic about the physical description of the disruption, has some limitations in accurately reproducing perihelion distances. By plotting the same distribution on a linear scale and as a difference between the observed and the predicted distributions, it becomes clear that there are two additional offsets at $q\sim0.7\au$ and $q\sim1\au$ where the model under-predicts the number of NEO detections (Fig.~\ref{fig:absolutediff_vs_q_17H25}). That is, there are systematically more NEOs on orbits for which perihelion distance coincides with the semimajor axes of Venus and the Earth, respectively, and the same trend is also apparent in Fig.~11 by \citet{2018Icar..312..181G} which presents an alternative approach to modeling the lack of NEOs at small $q$.

First we need to consider the possibility that the model's inability to predict enough NEO detections with $q~\sim a_{\rm planet}$ would be a modeling artifact. Given that we have no direct influence on the outcome of the fitting procedure---the debiased orbital model---the only alternative explanations are that the correction bias function and/or the input steady-state orbit distributions are erroneous. It is rather straightforward to rule out the possibility that the correction bias would be erroneous: despite the fact that the bias function has been carefully scrutinized, we could imagine an unlikely scenario where the detectability of Earth-approaching NEOs as observed from the Earth would have been estimated incorrectly. However, there is no conceivable reason why the detectability of NEOs with $q \sim a_{\rm Venus}$, as observed from the Earth, would also have been estimated incorrectly. Note that these excess NEOs are not necessarily detected close to the planet in question.

\begin{figure*}
    \centering
    \includegraphics[width=0.45\textwidth]{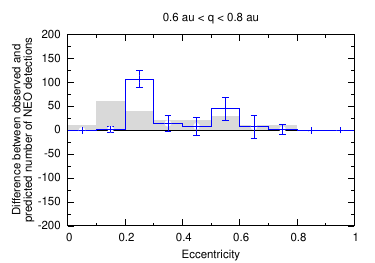}
    \includegraphics[width=0.45\textwidth]{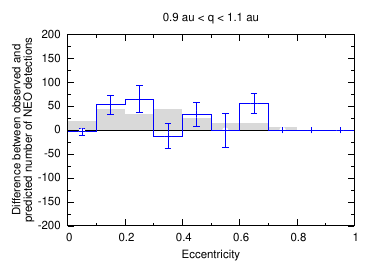}
    \includegraphics[width=0.45\textwidth]{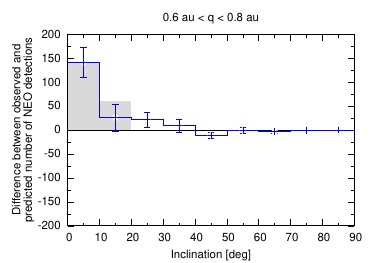}
    \includegraphics[width=0.45\textwidth]{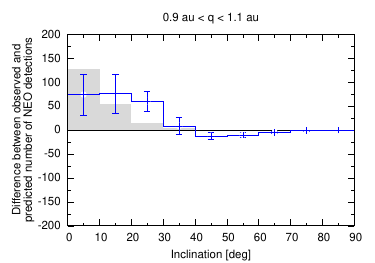}
    \caption{The difference (blue line) between observed and predicted number of NEO detections by CSS during the years 2005--2012 as a function of eccentricity $e$ (top panels) and inclination $i$ (bottom panels) for perihelion distances coinciding the semimajor axis of Venus (left panels) and the semimajor axis of the Earth (right panels). The model prediction assumes a super-catastrophic disruption when $q\sim0.076\au$ \protect \citep{2016Natur.530..303G}.
    The gray histograms show arbitrarily-normalized distributions of $e$ and $i$ of synthetic gravitational aggregates that in numerical simulations have undergone B-type tidal disruptions during encounters with the Earth or Venus.}
    \label{fig:absolutediff_vs_e_i_17H25}
\end{figure*}

The orbital integrations that were carried out to produce the steady-state orbit distributions took into account gravitational perturbations by all planets, and used a time step of 12 hours \citep{2018Icar..312..181G}. Only incorrectly modelled close encounters with terrestrial planets could change the orbit distribution so that the discrepancy is only apparent for orbits that have $q~\sim a_{\rm planet}$. In principle, a close encounter by an NEO with a very high encounter speed could go undetected and thus produce artifacts in the orbit distributions. There is no evidence for such artifacts in the orbit distributions, and it is not even clear that such an artifact would produce an offset in the correct direction. In addition, the excess detections are related to low-inclination and low-to-moderate-eccentricity orbits, that is, orbits that generally lead to slow encounter velocities (Fig.~\ref{fig:absolutediff_vs_e_i_17H25}), so an explanation based on undetected close encounters is not viable. 

\begin{figure}[h!]
  \centering
  \includegraphics[width=0.45\textwidth]{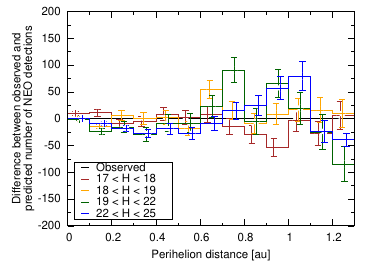}
  \caption{The difference between observed and predicted number of NEO detections by CSS during the years 2005--2012 as a function of perihelion distance $q$ separated into four different ranges in absolute magnitude $H$. The model prediction assumes a super-catastrophic disruption when $q\sim0.076\au$ \protect \citep{2016Natur.530..303G}. The excess detections at $q \sim 0.7\au$ and $q \sim 1\au$ correspond, in general, to smaller NEOs. See main text for caveats affecting interpretation.}
  \label{fig:absolutediff_vs_q_4Hbins}
\end{figure}

The excess detections in the \citet{2016Natur.530..303G} model primarily correspond to smaller NEOs with $18<H<22$ for those with $q \sim 0.7\au$ and $19<H<25$ for those with $q \sim 1\au$ (Fig.~\ref{fig:absolutediff_vs_q_4Hbins}). The largest NEOs considered by \citet{2016Natur.530..303G}, that is, those with $17<H<18$ do not show any evidence of excess detections. The breakdown of the excess detections into bins of $H$ are less certain than their bulk signature, and there are some caveats that need to be considered when interpreting the $H$ distributions of the excess detections. First, the fitting routine is trying to reproduce the observed distribution of NEO orbits and absolute magnitudes as accurately as possible, which implies that it will try to compensate for any shortcomings in the model's physical representation of the NEO population. That is, misleading compensation occurs, and we can only argue that some essential physics is missing from the model setup when there are too many (or too few) detections that can no longer be compensated for---which is exactly the case here with the excess detections. Hence the $H$ distribution of the excess detections, that the model cannot reproduce, will be a misleading representation of the $H$ distribution that would result if the missing physics would be accounted for. Second, low-eccentricity NEOs with $H>22$ are largely undetectable at $q<0.8\au$ (cf.\ Fig.~\ref{fig:absolutediff_vs_e_i_17H25}) and $q>1.2\au$. Third, the fitting by \citet{2016Natur.530..303G} was done using an extended maximum-likelihood scheme which aims to reproduce the total number of detections in addition to their distribution. Hence an excess in one part of the model may be counteracted with deficit in another. In summary, the excess detections preferentially correspond to small NEOs but the detailed $H$ distribution remains a topic of future studies.

Let us now assume that the excess detections correspond to fragments from tidal disruptions, and compare the expected orbits of those fragments to the orbits of the NEOs corresponding to the excess detections. Tidal disruptions have been classified by the amount of mass remaining in the disrupted body following its encounter with a planet: S-type encounters are extremely disruptive removing 90\% of the total mass whereas B-type disruptions remove 50-90\% of the total mass, and M-types remove less than 10\% \citep{1998Icar..134...47R}. S-type and B-type disruptions can thus generate a few or more large tidal-disruption fragments (compared to the parent body) and a significantly larger number of smaller fragments, whereas M-type disruptions only result in small fragments. While the details of the encounters such as spin and shape do matter, here we adopt the encounters that produce B-type disruptions for bodies with an average rotation period, and extract about 100 samples of progenitor orbits for close-enough and slow-enough encounters from published NEO orbit simulations \citep[][Zhang and Michel personal communication]{2010Icar..209..510N}. The disruption limits are scaled to a bulk density of $1.6\g\cm^{-3}$ and to a rotation period of 7~hr, both approximate averages for the NEO population \citep{2021pdss.data...10W}. The arbitrarily-normalized distributions of orbits leading to and immediately following B-type tidal disruptions are shown as the gray histograms in Figs.~\ref{fig:absolutediff_vs_q_17H25} and \ref{fig:absolutediff_vs_e_i_17H25}, and show an excellent agreement with the orbits corresponding to excess NEOs: objects that are most susceptible to tidal disruptions have low-to-moderate eccentricities and low inclinations. The lack of excess low-$e$ NEO detections with $0.6\au<q<0.8\au$ can be explained by accounting for the fact that NEOs with $e\lesssim0.2$ and $q<0.8\au$ never reach opposition as seen from the Earth, which makes them challenging to detect. That is, we cannot rule out tidal disruptions of NEOs with $e\lesssim0.2$ at Venus just based on an apparent lack of excess detections obtained from the Earth. We propose that the excess of low-$i$ Aten asteroids is at least partly explained as fragments from tidal disruptions \citep{2012ApJ...752..110M,2013ApJ...767L..18G}.

The fragments from recent tidal disruptions have small minimum orbital intersection distances (MOID) and slow speeds relative to the planet that caused the tidal disruption. Therefore, if tidal disruptions have occurred in the relatively recent past, we should expect to see an excess of small NEOs with slow relative speeds and close encounters when comparing to an orbital model that does not account for tidal disruptions. This is exactly what is seen in Figs.~5 (only NEOs detected by ATLAS) and 6 (all NEOs detected) in \citet{2021PSJ.....2...12H}, which compares NEO detections by ATLAS and other surveys to the model by \citet{2018Icar..312..181G}. Note that the normalization used makes it challenging to estimate the magnitude of the discrepancy.

\begin{figure*}
    \centering
    \includegraphics[width=\columnwidth]{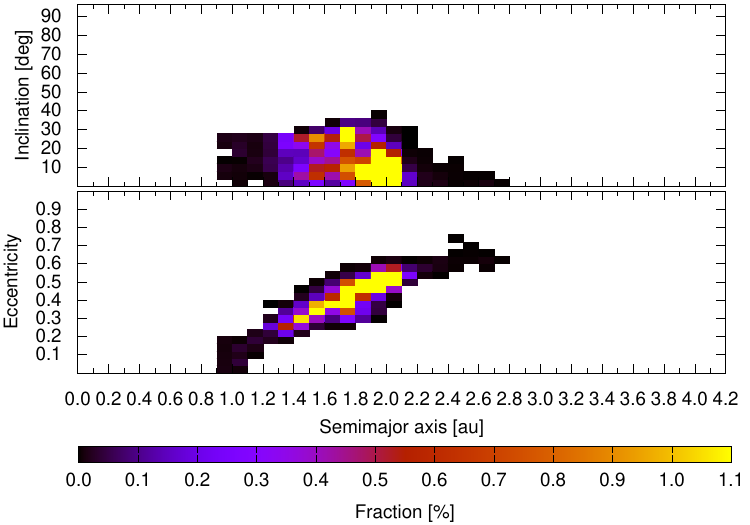}
    \includegraphics[width=\columnwidth]{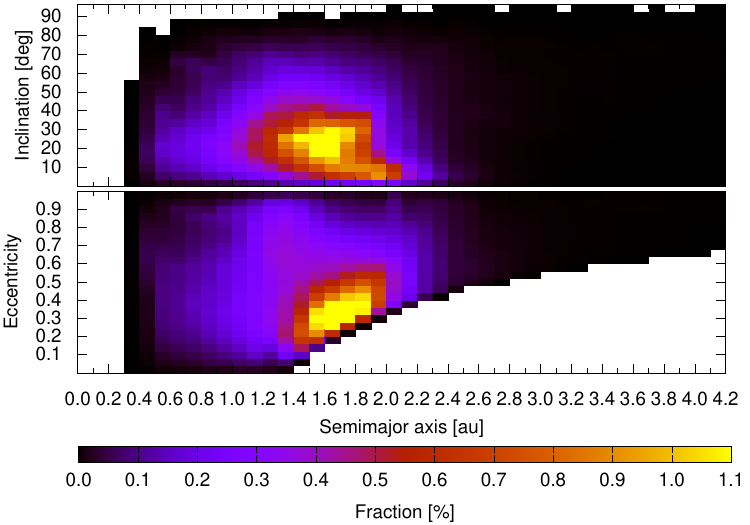}
    \caption{Examples of ensemble orbit distributions that could result from a large number of tidal disruptions of NEOs with $a<2\au$ and $i<25\deg$ during close encounters with the Earth $10\kyr$ after the disruption (left) and when all test asteroids have reached a sink (right). The assumption here is that the fragments are ejected at negligibly slow speeds relative to the disrupting parent body, which is corroborated by numerical simulations of tidal disruptions \citep{2014Icar..238..156S}, so only the orbital evolutions of the parent bodies are considered here.}
    \label{fig:orbitdistributions}
\end{figure*}

To further test the hypothesis of tidal disruptions being responsible for the excess detections, we generated orbit distributions corresponding to tidal disruptions at Venus and the Earth at different stages of their evolution and re-fitted the population models with these additional source regions for NEOs with $17<H<25$. The orbit distributions were derived by recording the evolution of the test asteroids used for the steady-state orbit distributions by \citet{2018Icar..312..181G} but selecting only those with orbital elements similar to the simulated gravitational aggregates that suffered tidal disruptions (gray histograms in Figs.~\ref{fig:absolutediff_vs_q_17H25} and \ref{fig:absolutediff_vs_e_i_17H25}). The time of entering the orbital space potentially leading to tidal disruptions also marked the starting point for recording their orbital evolution. Figure~\ref{fig:orbitdistributions} shows two examples of ensemble orbit distributions at different stages in their evolution resulting from tidal disruptions during an encounter with the Earth. The diffusion of the orbital elements over time is clearly visible, yet the location of the core of the distribution hardly changes from $10\kyr$ after the disruption until the time when all test asteroids have reached a sink, that is, a collision with a planet or the Sun, or an ejection from the inner Solar System due to a close encounter with a planet, typically Jupiter. The average lifetime for a test asteroid to reach a sink after a tidal disruption is $8.7\Myr$ whereas the 5th percentile is $0.03\Myr$ and the 95th percentile is $47\Myr$.

\begin{figure}[h!]
  \centering
  \includegraphics[width=0.45\textwidth]{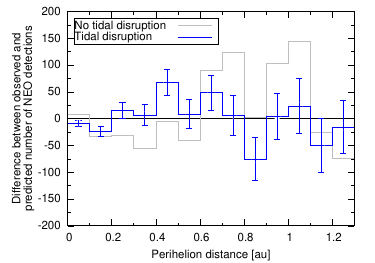}
  \caption{The difference between observed and predicted number of NEO detections by CSS during the years 2005--2012 as a function of perihelion distance $q$ when including orbit distributions that could result from tidal disruptions in the model (blue line). The model accounts for super-catastrophic disruptions by fitting for the parameters of a penalty function at small $q$ \protect \citep{2018Icar..312..181G}. The gray histogram is the one not accounting for tidal disruptions (Fig.~\ref{fig:absolutediff_vs_q_17H25}).}
  \label{fig:absolutediff_vs_q_17H25_tidaldisruptions}
\end{figure}

Since the focus here is on tidal disruptions occurring at relatively large $q$, we decided to use the modeling approach described by \citet{2018Icar..312..181G} who account for the super-catastrophic disruptions at small $q$ with a linear, two-parameter penalty function in the $(q,N)$ space, where $N=N(q)$ is the incremental number of NEO detections as a function of $q$. The chosen method improves the accuracy of the fit at small $q$ at the cost of making the interpretations somewhat less intuitive. The resulting $q$ distribution shows a significantly better agreement with the observed $q$ distribution for large $q$, and thus supports the hypothesis that tidal disruptions would be the explanation for the excess NEO detections (Fig.~\ref{fig:absolutediff_vs_q_17H25_tidaldisruptions}). 

An interesting feature arising from the new fit is the peak at $0.4\au<q<0.5\au$, which coincides with the semimajor axis of Mercury. Since tidal disruptions during Mercury encounters were not considered, the excess detections could be a signal of unaccounted tidal disruptions with Mercury. We note that Mercury encounters are not likely to lead to a significant rate of tidal disruptions, because the mass of Mercury is rather small, and the encounter speeds are typically large. The question will remain a topic of future studies given the limited statistics in the relevant part of the orbital space used in the present study as well as our current lack of knowledge about the mechanism(s) causing super-catastrophic disruptions---a major factor affecting the orbital distributions close to the Sun.

The fragments resulting from a tidal disruption will remain on planet-approaching orbits also for some time after a tidal-disruption event. Some fragments may therefore undergo further tidal disruption during subsequent close encounters with the planet, and thus increase the number of resulting fragments whereas some may impact the planet. There should therefore be at least an intermittent increase in the rate of close encounters and impacts with the planet following a tidal disruption (cf.\ Shoemaker-Levy 9). We estimated the increase in the long-term impact rate when accounting for tidal disruptions, and found that for $H<25$ the annual impact rate increases from 0.0012 \citep{2018Icar..312..181G} to 0.0018, or about 50\%, when using the same methodology for calculating the impact rate. 

In addition to increasing the rate of impacts with the Earth, fragments from tidal disruptions also increase the rate of impacts on nearby bodies. \citet{2018JGRE..123.2380W} describe the strong apex-to-antapex asymmetry of "cold spot" lunar craters that are only 0.023--2.3$\km$ in diameter and interpreted to be only 0.5--1$\Myr$ old. The size and asymmetry may be an indication of preferential formation by a population of projectiles with low relative speeds with respect to the Earth-Moon system that match the general properties of the fragments generated by tidal disruption.

Finally, as other mechanisms that lead to asteroid disruptions, also tidal disruptions produce dust and small meteoroids. The consequence of the fact that tidal disruptions happen close to planets is that the dust and small meteoroids will also remain on orbits that intersect that of Earth's for some time after the disruption, and should be detectable by meteor radars. We note that tidal disruptions are not necessarily one-off events, because close encounters can come in sequences of so-called resonant returns \citep{2003A&A...408.1179V}. Hence, either the parent body or its fragments---the latter formed in tidal disruptions during previous close encounters---may effectively produce a cascade of tidal-disruption events over an extensive period of time. On the other hand, the low-$i$ and low-$e$ of NEOs most prone to tidal disruption decreases the encounter speed, which, in turn, reduces the ionization and thus the radar detectability. In addition, the solar radiation pressure and frequent planetary encounters on such orbits diffuse the stream relatively fast until it becomes unidentifiable above the sporadic background. The Poynting-Robertson drag works on longer timescales, and reduces the heliocentric distance of the particles and circularizes the orbits. A detailed study of the longevity of circumsolar dust rings formed by tidal disruptions is left for future work, but they have been detected close to Mercury's and Venus' orbits \citep{2019ApJ...873L..16P,2023PSJ.....4...33P}, and, to the best of our knowledge, a formation scenario including tidal disruption of NEOs has not been considered to date.

\section{Conclusions}
\label{sec:conclusions}

We have shown that the tidal disruption of asteroids during close encounters with the Earth and Venus is an observational fact, and potentially solves a number of open issues that are linked to NEOs on orbits that are either similar or tangential to those of the terrestrial planets. The discovery expands on the work by \citet{2010Natur.463..331B} who proposed that close encounters with the Earth, that are more distant than those considered here, refresh the surfaces of asteroids.

We speculate that, in the future, it will be possible to make a statistically-significant identification of the much weaker signal from tidal disruptions during Mercury encounters. Such an identification requires better statistics of NEOs with $q \sim a_{\rm Mercury}\sim0.4\au$ and also a reasonably accurate model of the super-catastrophic disruptions at small $q$.

We stress that these results do not suggest that tidal disruption during close planetary encounters would be the primary mechanism destroying gravitational aggregates in the inner Solar System. Moreover, here we report an overabundance of NEO detections, which implies a generation of more observable NEOs, not fewer. The benefit compared to other disruption mechanisms, such as rotational disruption, is that the frequency of planetary encounters and the subsequent dynamical evolution of the fragments is well understood, which allows for testing the susceptibility of asteroids for tidal disruption on the population level.

\begin{acknowledgments}
We thank the anonymous referee whose suggestions improved the manuscript. MG was partly funded by the Swedish Research Council's award number 2022-04615.
\end{acknowledgments}

\bibliography{2023_neotidal}{}
\bibliographystyle{aasjournal}



\end{document}